\documentclass[nonacm,sigconf]{acmart}
\usepackage{balance}
\usepackage{array}
\AtBeginDocument{%
  }

\usepackage{subfig}
\usepackage{threeparttable}

\begin{document}

\title{Acceptance and Responsibility Attribution Toward Autonomous Ambulances Across Driving Conditions: A Preliminary Survey of Fire Service Personnel}

\author{Hirotake Yamazoe}
\affiliation{%
  \institution{University of Hyogo}
  \city{Himeji}
  \state{Hyogo}
  \country{Japan}
}
\email{yamazoe@eng.u-hyogo.ac.jp}
\orcid{0000-0001-9995-4346}

\author{Shingo Yamaguchi}
\affiliation{%
  \institution{Himeji Fire Bureau}
  \city{Himeji}
  \state{Hyogo}
  \country{Japan}
}

\renewcommand{\shortauthors}{Yamazoe and Yamaguchi}


\begin{abstract}
Traffic incidents involving ambulances and fire engines during emergency-response driving can have serious consequences. Although autonomous driving technologies may reduce workload and improve safety, their introduction into emergency services requires consideration of first responders' trust, willingness to ride (WTR), and perceptions of responsibility.
We surveyed 34 fire service personnel from a single fire station in Himeji, Japan, to examine their evaluations of autonomous ambulances and responsibility attribution under non-emergency and emergency driving conditions. Emergency driving was associated with lower perceived usefulness for improving safety, trust, and WTR, as well as greater anxiety. In contrast, responsibility attribution remained largely stable, with 30 participants ($88.2\%$) selecting the same primary responsible party across conditions and the onboard crew most frequently selected in both. Among these 30 participants, 26 ($86.7\%$) also assigned the same responsibility share.
Open-ended responses revealed concerns about system reliability, the complexity of emergency driving, and the need for human judgment, as well as considerations regarding operator and organizational responsibility. These findings suggest that acceptance of autonomous ambulances is sensitive to driving conditions, while responsibility judgments may also reflect perceived professional roles and responsibilities.
\end{abstract}




\maketitle

\section{Introduction}
Traffic incidents involving ambulances and fire engines during emergency-response driving can have serious consequences, including delays in patient care and loss of public trust in emergency services. 
Previous studies have investigated the occurrence, characteristics and risk factors of ambulance crashes~\cite{Delavary2023,mulvihill2025extent,Norii2024},
while other research has explored approaches to improving emergency-driving safety, including driver feedback, simulator-based training, and driving assistance systems~\cite{levick2005optimal,simeonov2021evaluation,prohn2022traffic}.
As part of this broader effort, our group has developed a VR emergency-response driving simulator in collaboration with the Himeji Fire Bureau to support accident-prevention education through driving-behavior analysis and replay-based debriefing~\cite{sudou2024preliminary,te2026yamazoe}. Although the simulator was not used in the present questionnaire study, it could serve as a platform for future evaluations of emerging driving-support and automated driving technologies for emergency vehicles.

Advances in automated driving may enable its application to emergency vehicles, potentially reducing workload and improving safety. However, introducing such technologies requires consideration of onboard personnel's roles, the delegation of driving authority, and responsibility for crashes or operational failures, in addition to technical performance.

Previous research has examined the acceptance of automated vehicles using technology-acceptance frameworks and has identified perceived safety, anxiety, trust, and perceived risk as relevant factors~\cite{osswald2012predicting,hewitt2019assessing,zhang2019roles}. Studies of autonomous ambulances have primarily focused on potential patients~\cite{winter2018patient,zarkeshev2019ready,zarkeshev2020patients}, while research involving first responders has mainly addressed their attitudes toward and interactions with non-emergency connected and automated vehicles~\cite{almaskati2025review,xu2025understanding,liu2023first}. Responsibility attribution studies have further shown that perceived responsibility does not necessarily correspond to actual control or the ability to prevent a crash~\cite{beckers2022drivers,zhai2024not}. However, how fire service personnel attribute responsibility for crashes involving autonomous ambulances remains insufficiently understood.

In an autonomous ambulance, onboard personnel may provide emergency medical care while also being expected to monitor automated driving operations or intervene when necessary. 
This combination of roles raises questions not only about trust and willingness to use the vehicle but also about the authority retained by onboard personnel, professional duties, and responsibility when a crash occurs. In the present study, responsibility attribution refers to respondents' perceived allocation of operational responsibility rather than a determination of legal liability.

Emergency-response driving also differs from non-emergency driving in terms of urgency, traffic-rule exemptions, and operational risk. These contextual differences may affect acceptance-related evaluations, such as perceived usefulness for improving safety, anxiety, trust, and WTR. Responsibility attribution may also reflect assumptions concerning professional duties, institutional roles, and the authority to monitor or intervene.

In our previous report, we compared fire service personnel's evaluations of autonomous ambulances under non-emergency and emergency driving conditions and reported the distribution of the parties selected as primarily responsible for a crash under each condition~\cite{yamazoe2026tbsa}. 
The results suggested that acceptance-related evaluations varied across driving conditions, while the onboard crew remained the most frequently selected primary responsible party under both conditions.

Using the same survey dataset, the present study extends the previous report by examining within-participant changes in the selection of the primary responsible party, comparing the proportions of responsibility assigned across driving conditions, and analyzing optional open-ended responses.
The study addresses the following research questions:

{
\setlength{\leftmargini}{20pt}
\begin{itemize}
\item \textbf{RQ1:} Which party is selected as primarily responsible for a crash under non-emergency and emergency driving conditions, and how stable is this selection within participants?

\item \textbf{RQ2:} 
What proportion of total responsibility do participants assign to the selected primary responsible party, and how does this proportion vary across driving conditions and responsible parties?

\item \textbf{RQ3:} 
What concerns and considerations regarding automated driving technology, emergency driving, and responsibility attribution are reflected in the optional open-ended responses, and how do they relate to the quantitative findings?
\end{itemize}
}

Together, these analyses provide a more detailed understanding of how fire service personnel attribute responsibility for autonomous ambulance crashes, drawing on within-participant comparisons and open-ended responses.
The study also examines how responsibility attribution and acceptance-related evaluations vary across non-emergency and emergency driving conditions.

\section{Related Work}

\subsection{Acceptance of Automated Vehicles and Autonomous Ambulances}
Previous studies have examined the acceptance of automated driving technologies using various technology acceptance models. 
In the automotive domain, the Car Technology Acceptance Model (CTAM) emphasized perceived safety and anxiety, while the Autonomous Vehicle Acceptance Model (AVAM) extended technology acceptance frameworks to different levels of vehicle automation~\cite{osswald2012predicting,hewitt2019assessing}.
Other studies have identified trust and perceived risk as important factors in the acceptance of automated vehicles~\cite{zhang2019roles,nordhoff2021perceived}. 
These studies provide relevant background for the acceptance-related measures examined in the present study, although we do not test or compare the models themselves.

Studies of autonomous ambulances have mainly examined potential patients. Winter et al. found lower willingness to ride in an autonomously operated ambulance than in a conventionally driven ambulance, with affect helping to explain this difference~\cite{winter2018patient}. Studies by Zarkeshev and Csiszár also reported limited readiness to use autonomous ambulances and emphasized the need for public information and education~\cite{zarkeshev2019ready,zarkeshev2020patients}. Research involving first responders has reported relatively low trust in connected and automated vehicles and highlighted the importance of familiarity, training, and standardized operational protocols~\cite{liu2023first,xu2025understanding,almaskati2025review}. However, these studies have primarily addressed first responders' attitudes toward and interactions with CAVs in general, rather than their perceptions of automated emergency vehicles used in their own operations.

\subsection{Responsibility Attribution in Automated Driving}
Automated driving complicates responsibility attribution because control and decision-making may be distributed across users, automated systems, manufacturers, and operating organizations. 
Previous studies have highlighted challenges in attributing responsibility when users have limited control over automated systems~\cite{hevelke2015responsibility,elish2019moral}. In particular, Elish introduced the concept of a “moral crumple zone,” describing how human operators may be disproportionately blamed for failures in automated systems despite their limited control~\cite{elish2019moral}.
Empirical studies have also shown that humans may be blamed for crashes even when automated systems share control or limit their ability to intervene~\cite{awad2020blaming,beckers2022drivers}.

Zhai et al. found that increasing levels of vehicle automation were associated with reduced responsibility attributed to humans, partly explained by perceived human controllability~\cite{zhai2023human}. Nevertheless, perceived human responsibility can persist even at higher levels of automation. Occupants of fully automated vehicles have been assigned responsibility despite lacking direct control, partly because they were expected to foresee possible harm~\cite{zhai2024not}. Meanwhile, Li et al. found that participants attributed greater responsibility to manufacturers and the government when an autonomous vehicle caused a crash than when a human driver caused the crash~\cite{li2016trolley}. Increasing automation may also shift perceived responsibility toward manufacturers, operators, and other corporate or institutional entities~\cite{simmler2024responsibility}. More recently, a study analyzing social media comments identified automated driving systems, human drivers, and vehicle-related publicity as targets of public blame for automated vehicle crashes~\cite{sun2025blame}. These studies suggest that responsibility attribution is not determined solely by physical control, but may also reflect the perceived foreseeability of harm and considerations concerning different stakeholders.

Autonomous ambulances present a distinctive case because onboard personnel may be expected to provide emergency medical care while also supervising automated driving operations or intervening when necessary. To our knowledge, few studies have examined how fire service personnel attribute responsibility for crashes involving autonomous ambulances or whether their responsibility judgments differ between non-emergency and emergency driving conditions.

\section{Method}
\subsection{Participants and Procedure}
To explore fire service personnel's evaluations of autonomous ambulances and their attribution of responsibility for crashes, we conducted an online questionnaire survey with personnel from a single fire station in Himeji. A total of 34 personnel participated in the survey ($N = 34$).

For the purposes of the survey, \textit{autonomous driving} was defined as a high level of driving automation in which the automated driving system performs the driving task and onboard personnel are not required to operate the steering wheel or pedals. Participants were instructed to assume a near-future scenario in which autonomous vehicles, including autonomous ambulances, were widely available and commonly used.

The questionnaire was administered using Microsoft Forms.
Participants first reviewed the objectives of the study,
the assumed future scenario, and the definition of autonomous driving. Participation was voluntary, and submission of the completed questionnaire was considered to indicate informed consent to participate in the study.

\subsection{Measures}
The questionnaire covered the following six areas:
(1) experience in emergency vehicle operations;
(2) experience with and familiarity with autonomous vehicles and advanced driver-assistance systems (ADAS);
(3) willingness to ride (WTR) and trust regarding autonomous vehicles in general;
(4) the perceived necessity and usefulness of ambulance automation;
(5) perceived usefulness for improving safety, anxiety, trust, and WTR regarding autonomous ambulances when serving as onboard personnel under non-emergency and emergency driving conditions;
and (6) responsibility attribution in the event of a crash.

Acceptance-related evaluations were measured using seven-point rating scales. Optional free-response fields allowed participants to explain the reasons for their evaluations and to describe concerns or other considerations. For each evaluation item involving both driving conditions, the non-emergency driving condition was presented before the emergency driving condition. The questionnaire items and response formats are summarized in Table~\ref{tab:questionnaire}.

\begin{table}[tb]
    \centering
    \caption{Details of Questionnaire Items}
    \vspace{-3mm}
    \label{tab:questionnaire}
  \scalebox{0.90}{
    \begin{tabular}{>{\raggedright\arraybackslash}p{1.05\columnwidth}}
       \hline
       \textbf{(1) Experience with emergency vehicle operations} \\
        - Years of experience in emergency vehicle operations, including non-driving roles \\
        - Frequency of emergency vehicle dispatches during the previous year \\
        - Proportion of recent dispatches involving ambulances \\
	    -  Proportion of recent dispatches in which the participant served as the driver \\
        - Proportion of ambulance dispatches across the participant's overall emergency vehicle experience \\   
       \hline
       \textbf{(2) Experience and familiarity with autonomous vehicles and ADAS} \\
        - Previous experience riding in an autonomous vehicle (e.g., robotaxi, autonomous bus, or similar pilot services) \\
        - Names of the autonomous services used (open-ended if applicable) \\
        - Experience using various ADAS features \\
        - Frequency of ADAS use \\
       \hline
       \textbf{(3) Willingness to ride (WTR) and trust in general autonomous vehicles (GAVs)} \\
	- WTR in GAVs in daily life \\
	- Perceived trust in GAVs \\
       \hline
	\textbf{(4) Perceived necessity and usefulness of automating ambulances}  \\
	- Necessity of automating ambulances in the future \\
	- Usefulness of autonomous ambulances (AAs) for reducing operational workload \\
	- Perceived usefulness of AAs for improving safety under non-emergency conditions\\
	- Perceived usefulness of AAs for improving safety under emergency driving conditions\\
       \hline
	\textbf{(5) WTR in, trust in, and anxiety about autonomous ambulances when serving as onboard personnel} \\
	- WTR in an AA under non-emergency conditions \\
	- WTR in an AA under emergency driving conditions \\
	- Perceived trust in an AA under non-emergency conditions \\
	- Perceived trust in an AA under emergency driving conditions \\
	- Anxiety about an AA under non-emergency conditions \\
	- Anxiety about an AA under emergency driving conditions \\
	\hline
	\textbf{(6) Operational responsibility attribution in the event of a crash} \\
	- Party considered most responsible for a crash under non-emergency driving conditions \\
	- Party considered most responsible for a crash under emergency driving conditions \\
	- Proportion of total responsibility assigned to the party selected as most responsible under each condition (0–10, with 10 representing 100\%)\\
    \textit{Options: autonomous driving system/manufacturer, crew's organization (fire department), onboard crew, or other} \\
	\hline
    \end{tabular}
	}
\end{table}

\subsection{Quantitative Analysis}
We used paired $t$-tests to compare acceptance-related evaluations under non-emergency and emergency driving conditions, as reported in our previous study~\cite{yamazoe2026tbsa}.
All statistical tests were two-sided and conducted using R version 4.5.1.

Within-participant changes in the selection of the primary responsible party were examined by comparing responses under the two driving conditions. The frequencies and directions of changes were summarized descriptively. Responsibility shares were compared between conditions for participants who selected the same primary responsible party under both conditions. We also counted the number of participants who assigned identical responsibility shares under both conditions.

\subsection{Analysis of Open-Ended Responses}
We conducted a qualitative content analysis of the responses to five optional open-ended questions.

Generative AI (ChatGPT 5.6 Sol) was used as a preprocessing aid to organize candidate codes and support the development of an initial codebook. The codebook was iteratively reviewed and refined based on the original responses and the context of the corresponding questions. The final codebook consisted of six codes, as defined in Table~\ref{tab:qualitative_indicators}. The first author subsequently coded all responses manually using the final codebook and made all final decisions regarding code assignment.

Multiple codes could be assigned to a response when it contained more than one relevant consideration, whereas repeated instances of the same code within a response were counted only once. For descriptive comparisons with the corresponding quantitative questionnaire responses, we counted the number of participants who mentioned each code at least once across the five questions.

\begin{table}[tb]
\caption{Codebook for the analysis of open-ended responses.}
\label{tab:qualitative_indicators}
    \vspace{-3mm}
\centering
  \scalebox{0.90}{
\begin{tabular}{p{\columnwidth}}
\hline
\textbf{(a) Current distrust/anxiety}\\
Expressions of current distrust, anxiety, or reluctance toward the use of automated driving. \\
\textbf{(b) Emergency-context concern}\\
Concerns regarding the complexity of emergency driving, unpredictable behavior of surrounding road users, or the need for human judgment and experience. \\
\textbf{(c) System capability concern}\\
Concerns about whether the automated system can reliably recognize relevant situations and operate as required. \\
\textbf{(d) Operator responsibility}\\
References to the responsibility of onboard personnel or driver-related factors in the context of responsibility attribution. \\
\textbf{(e) Broader conditional responsibility}\\
References to organizational, system/manufacturer, shared, or condition-dependent responsibility. \\
\textbf{(f) Responsibility ambiguity/control concern}\\
References to ambiguity in responsibility attribution or the relationship between responsibility and system use or control authority.
\\
\hline
\end{tabular}
}
\end{table}

\section{Results}
\subsection{Participant Characteristics and Acceptance-Related Evaluations}
As reported in our previous study~\cite{yamazoe2026tbsa}, participants had 0--25 years of emergency vehicle experience, and 25 ($73.5\%$) had 6--20 years. 
Only one had used a high-level automated driving service (Waymo), while 31 ($91.2\%$) had used ADAS. However, 22 ($64.7\%$) reported ADAS use at or below the scale midpoint.

Mean WTR and trust ratings for general autonomous vehicles (GAVs) were 3.88 ($SD = 1.57$) and 3.32 ($SD = 1.45$), respectively. The perceived necessity of autonomous ambulances was rated at 3.29 ($SD = 1.75$), while usefulness for reducing workload was rated at 4.50 ($SD = 1.75$).

\begin{figure}[tb]
\centering
    \includegraphics[width=0.8\linewidth]{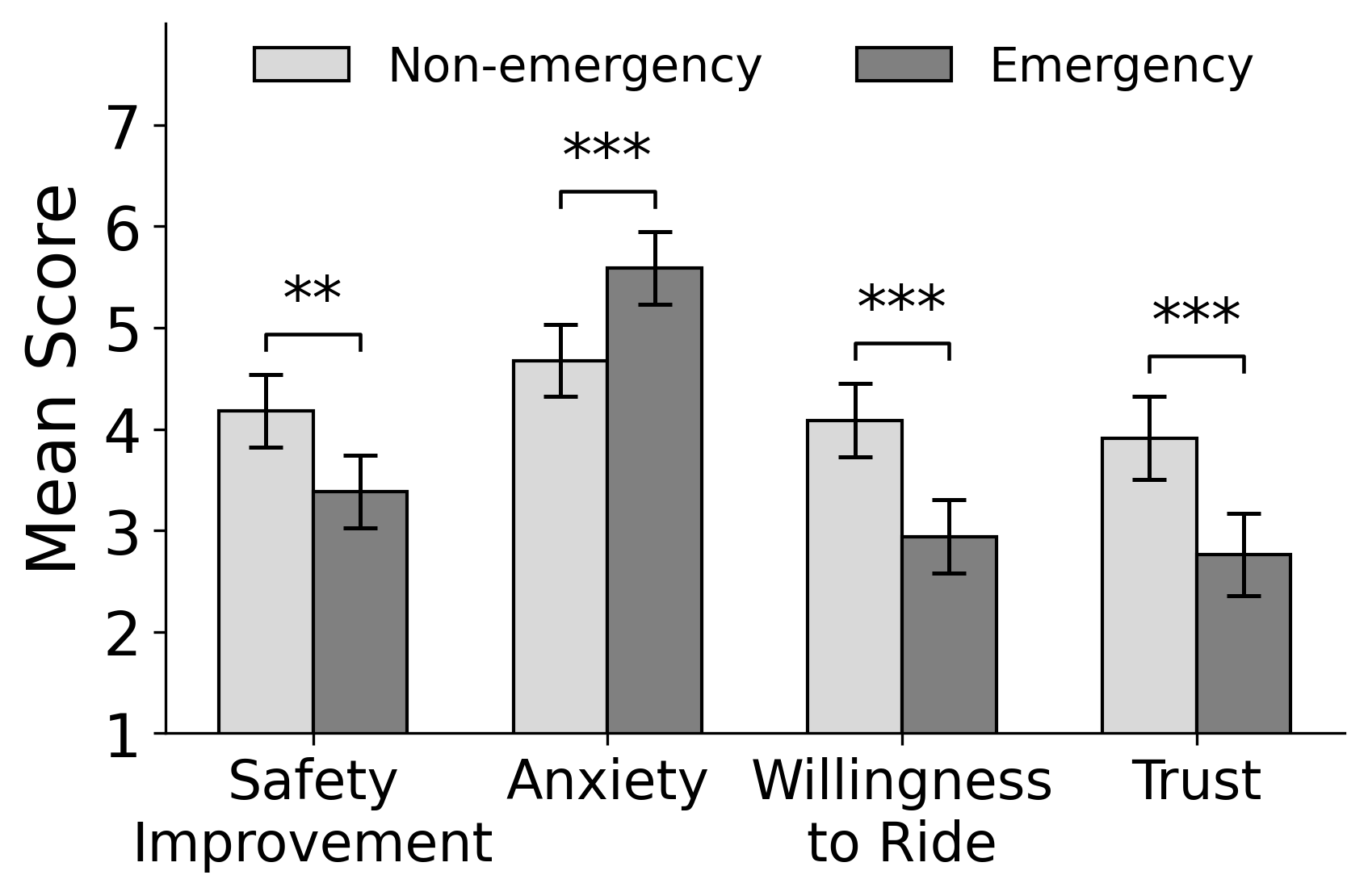}
    \vspace{-3mm}
    \caption{Perceived usefulness for improving safety, anxiety, WTR, and trust under non-emergency and emergency driving conditions. Error bars represent within-subject 95\% confidence intervals calculated using the Loftus--Masson method. $^{*}p < .05$ $^{**}p < .01$ $^{***}p < .001$ .}
    \Description{Perceived usefulness for improving safety, anxiety, WTR, and trust under non-emergency and emergency driving conditions.}
    \label{fig:difference}
\end{figure}

Evaluations of autonomous ambulances differed between non-emergency (AA-NE) and emergency (AA-E) driving conditions. Figure~\ref{fig:difference} shows the ratings of perceived usefulness for improving safety, anxiety, WTR, and trust under the two conditions. Paired t-tests indicated lower perceived usefulness for improving safety ($t(33) = -3.189$, $p = .0031$, $d_z=-.547$), WTR ($t(33) = -4.584$, $p < .001$, $d_z=-.786$), and trust ($t(33) = -4.044$, $p < .001$, $d_z=-.694$) under AA-E than under AA-NE, whereas anxiety was higher under AA-E ($t(33) = 3.684$, $p < .001$, $d_z=.632$). Overall, emergency driving was associated with less favorable acceptance-related evaluations than non-emergency driving.

\subsection{Responsibility Attribution Across Driving Conditions}
\begin{figure}[tb]
    \centering
    \includegraphics[width=\linewidth]{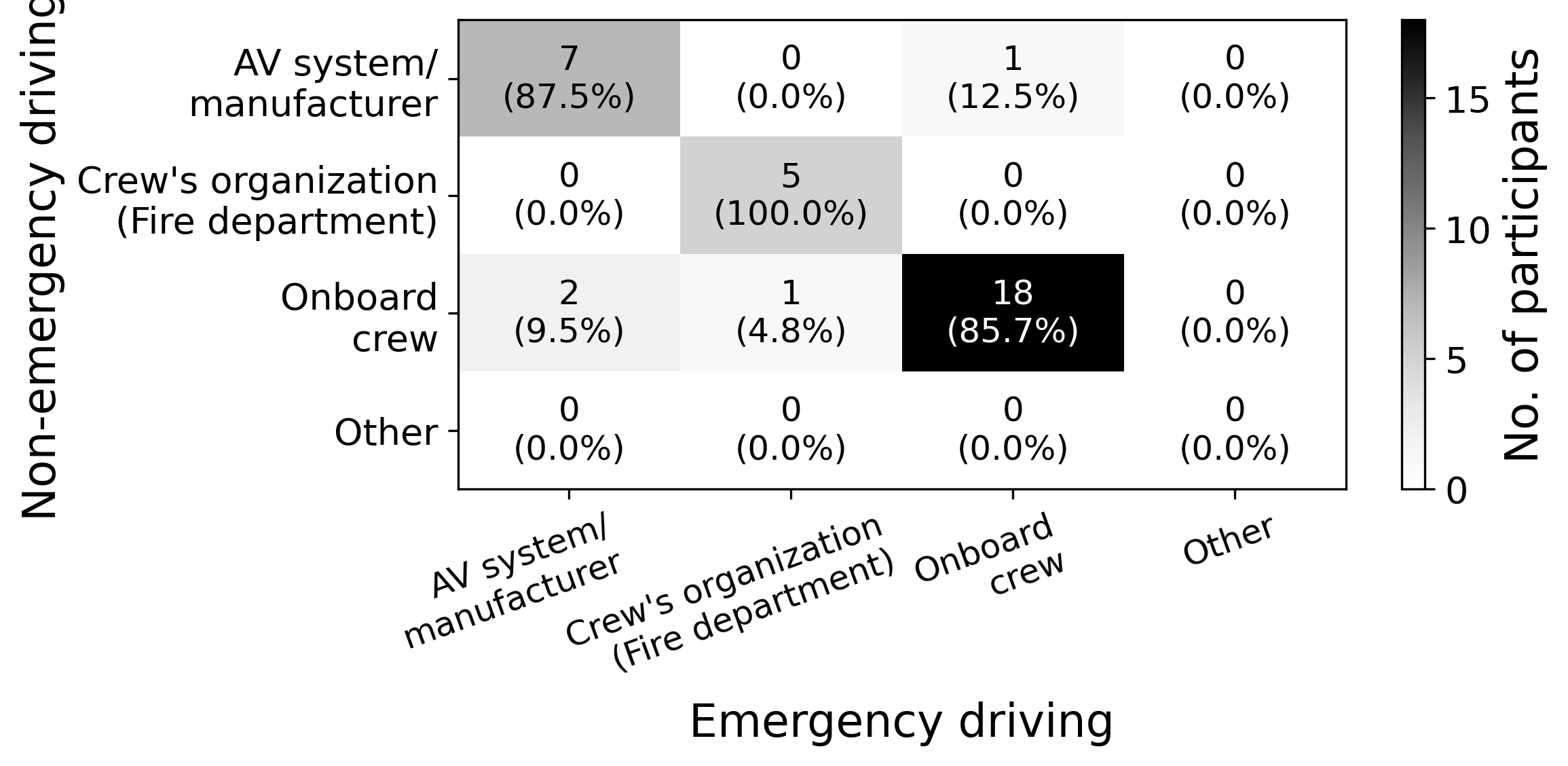}
    \vspace{-8mm}
    \caption{Party considered most responsible for a crash 
    under non-emergency and emergency driving conditions. Cell values indicate the number of participants, with row percentages shown in parentheses.}
    \Description{Party considered most responsible for a crash under non-emergency and emergency driving conditions.}
    \label{fig:responsibility}
\end{figure}

Figure~\ref{fig:responsibility}  shows the parties selected as primarily responsible for a crash under non-emergency and emergency driving conditions. Under non-emergency driving, 21 participants ($61.8\%$) selected the onboard crew, eight ($23.5\%$) selected the autonomous driving system/manufacturer, and five ($14.7\%$) selected the crew's organization. Under emergency driving, the corresponding numbers were 19 ($55.9\%$), nine ($26.5\%$), and six ($17.6\%$), respectively.

Most participants (30 of 34, $88.2\%$) selected the same primary responsible party under both conditions. Of the four participants who changed their selection, two shifted from the onboard crew to the autonomous driving system/manufacturer, one from the onboard crew to the crew's organization, and one from the autonomous driving system/manufacturer to the onboard crew.

Table~\ref{tab:responsibility_share} summarizes the responsibility shares assigned to the primary responsible party among the 30 participants who selected the same party under both conditions. Of these participants, 26 ($86.7\%$) assigned exactly the same responsibility share under both conditions.


\begin{table}[t]
\centering
\caption{Responsibility shares assigned to the same primary
responsible party across driving conditions.}
    \vspace{-3mm}
\label{tab:responsibility_share}
\begin{tabular}{lccc}
\toprule
Primary responsible party
    & $n$ & Non-emergency & Emergency \\
\midrule
Onboard crew
    & 18 & 8.44 (1.15) & 8.56 (1.04) \\
Crew's organization
    & 5  & 6.80 (1.10) & 6.80 (1.10) \\
ADS/manufacturer
    & 7  & 7.86 (1.77) & 7.14 (1.86) \\
\bottomrule
\end{tabular}
\smallskip
\footnotesize
\raggedright
\textit{Note.} Values are mean (SD).
Total responsibility was defined as 10 (100\%).\\
Only participants who selected the same primary responsible
party under both conditions were included ($n = 30$).
\end{table}

\subsection{Analysis of Open-Ended Responses}

Table~\ref{tab:open_ended_summary} summarizes the code frequencies. Current distrust/anxiety was most frequent (19 of 34 participants), followed by emergency-context (10) and system capability concerns (9).

Responsibility-related comments were less frequent, with four participants (11.8\%) mentioning each of the three responsibility-related codes. All four participants whose responses were coded as operator responsibility were among the 18 participants who consistently selected the onboard crew as the primary responsible party under both conditions. In contrast, participants who mentioned broader or conditional responsibility selected different types of parties as primarily responsible, including the onboard crew, their organization, and the autonomous driving system/manufacturer.

For example, one participant emphasized the driver's responsibility under both driving conditions while acknowledging that organizational management and unexpected system failures could also be relevant. Another participant stated that they would not use automated driving if the driver could be held responsible for a crash occurring during automated operation.

\begin{table}[t]
\centering
\caption{Summary of open-ended responses ($N=34$).}
    \vspace{-3mm}
\label{tab:open_ended_summary}
\begin{tabular}{lr}
\toprule
\textbf{Code} & \textbf{Participants, $n$ (\%)} \\
\midrule
Current distrust/anxiety              & 19 (55.9) \\
Emergency-context concern             & 10 (29.4) \\
System capability concern             &  9 (26.5) \\
\midrule
Operator responsibility               &  4 (11.8) \\
Broader/conditional responsibility    &  4 (11.8) \\
Responsibility ambiguity/control concern & 4 (11.8) \\
\bottomrule
\end{tabular}
\begin{flushleft}
\footnotesize
\end{flushleft}
\end{table}

\section{Discussion}
\subsection{Acceptance of Autonomous Ambulances Across Driving Conditions}
Although participants recognized the potential usefulness of autonomous ambulances for reducing operational workload, their acceptance-related evaluations were less favorable under emergency driving conditions. 
Specifically, perceived usefulness for improving safety, trust, and WTR were lower, whereas anxiety was higher, under emergency driving conditions than under non-emergency conditions. 
These findings suggest that acceptance-related evaluations may depend on the operational context and that evaluations under non-emergency conditions may not directly extend to emergency operations. This is consistent with previous research identifying perceived safety, trust, and risk as important factors in automated vehicle acceptance~\cite{nordhoff2021perceived,zhang2019roles}.

The open-ended responses provided additional context for these findings. Participants expressed concerns about the complexity of emergency driving, unpredictable behavior of surrounding road users, and the need for human judgment and experience. They also questioned whether automated systems could reliably recognize relevant situations and operate as required. These responses highlight concerns regarding both the suitability of automated driving for complex emergency operations and the reliability of the automated system itself.

\subsection{Responsibility Attribution Across Driving Conditions}
In contrast to the changes in acceptance-related evaluations, 
responsibility attribution remained largely stable across driving conditions. 
Most participants selected the same primary responsible party under both conditions, and most of these participants also assigned the same responsibility share.
Notably, the onboard crew was the most frequently selected primary responsible party, even though participants were instructed to assume a high level of driving automation without the need for manual driving operations.

The open-ended responses provided additional insights into these judgments. 
All four participants whose responses were coded as operator responsibility selected the onboard crew as the primary responsible party under both conditions. 
In contrast, participants who mentioned broader or conditional responsibility selected different types of parties as primarily responsible. These findings suggest that selecting a single primary responsible party may not fully capture participants' views on how responsibility should be distributed among onboard personnel, organizations, and system providers.

Previous studies have reported that human occupants or drivers may be held responsible for crashes even when automation limits their ability to intervene~\cite{beckers2022drivers,zhai2024not}. 
The present findings suggest that similar expectations may exist among fire service personnel evaluating autonomous ambulances. 
In addition, one participant expressed reluctance to use automated driving if the driver would be held responsible for a crash during automated operation, suggesting that responsibility concerns may also be relevant to acceptance. 
However, because the questionnaire did not specify crash circumstances or opportunities for human intervention, further research is needed to examine how these factors influence responsibility attribution and acceptance-related evaluations.

\subsection{Future Directions}
Future studies should investigate how video-, VR-, or simulator-based experiences with autonomous ambulance systems influence acceptance-related evaluations and responsibility attribution. Such experiences could help fire service personnel better understand the capabilities and limitations of automated driving under non-emergency and emergency conditions.

Future research should also examine how specific crash scenarios, system capabilities, opportunities for human intervention, and supervisory roles influence responsibility attribution. 
These investigations could help clarify how expectations concerning professional roles and control authority shape responsibility judgments and inform discussions about the roles and responsibilities of onboard personnel, fire departments, and system providers.

\subsection{Limitations}
This study has several limitations. First, it was a preliminary investigation involving 34 personnel from a single fire station, which limits the generalizability of the findings. Second, most participants lacked direct experience with high-level automated vehicles and evaluated hypothetical future scenarios. Their evaluations may differ when interacting with actual or simulated autonomous ambulance systems.

Third, the questionnaire did not specify detailed crash scenarios, system capabilities, or opportunities for human intervention. These factors may influence responsibility attribution, and the observed stability across driving conditions should therefore be interpreted cautiously. 
Although the questionnaire explicitly referred to operational responsibility, the term was not further defined, and participants may have interpreted it differently. 
Furthermore, participants were asked to select only the primary responsible party and indicate its responsibility share, without specifying how the remaining responsibility was distributed among other parties. 
Analyses of responsibility shares were descriptive and limited to participants who selected the same primary responsible party under both conditions. 
Several acceptance-related constructs were also assessed using single-item measures.

Finally, the open-ended questions were optional, and all responses were coded by a single researcher. The qualitative findings should therefore be interpreted cautiously, particularly given the limited number of responsibility-related comments.

\section{Conclusion}
This study examined fire service personnel's acceptance-related evaluations and responsibility attribution toward autonomous ambulances under non-emergency and emergency driving conditions. Emergency driving was associated with less favorable acceptance-related evaluations, whereas responsibility attribution remained largely stable, with the onboard crew most frequently selected as the primary responsible party.

Open-ended responses raised concerns about system reliability, the complexity of emergency driving, and human judgment.
These responses also suggested that selecting a single party as primarily responsible may not fully reflect participants' views on how responsibility should be distributed.

These findings highlight the importance of considering operational context and perceived responsibilities when evaluating autonomous ambulances. Future research should examine how system experience, human intervention opportunities, and supervisory roles influence these evaluations.

\begin{acks}
This research was supported in part by JSPS KAKENHI Grant JP23K26082.
\end{acks}

\bibliographystyle{ACM-Reference-Format}
\bibliography{sample-base}

\end{document}